# Large anomalous Hall effect and anisotropic magnetoresistance in intrinsic nanoscale spin-valve-type structure of an antiferromagnet

Dong Gun Oh[1,2], Jong Hyuk Kim[1,2], Mi Kyung Kim[1], Ki Won Jeong[1], Hyun Jun Shin[1], Jae Min Hong[1], Jin Seok Kim[1], Kyungsun Moon[1], Nara Lee[1,*], and Young Jai Choi[1,*]

A spin valve is a prototype of spin-based electronic devices found on ferromagnets, in which an antiferromagnet plays a supporting role. Recent findings in antiferromagnetic spintronics show that an antiferromagnetic order in single-phase materials solely governs dynamic transport, and antiferromagnets are considered promising candidates for spintronic technology. In this work, we demonstrated antiferromagnet-based spintronic functionality on an itinerant Ising antiferromagnet of $Ca_{0.9}Sr_{0.1}Co_2As_2$ by integrating nanoscale spin-valve-type structure and investigating anisotropic magnetic properties driven by spin-flips. Multiple stacks of 1 nm thick spin-valve-like unit are intrinsically embedded in the antiferromagnetic spin structure. In the presence of a rotating magnetic field, a new type of the spin-valve-like operation was observed for large anomalous Hall conductivity and anisotropic magnetoresistance, whose effects are maximized above the spin-flip transition. In addition, a joint experimental and theoretical study provides an efficient tool to read out various spin states, which scheme can be useful for implementing extensive spintronic applications.

[1]Department of Physics, Yonsei University, Seoul 03722, Korea. [2]These authors contributed equally: D. G. Oh, J. H. Kim. *email: eland@yonsei.ac.kr; phylove@yonsei.ac.kr

## INTRODUCTION

Antiferromagnetism was discovered in the 1930s as an elementary form of spin order after ferromagnetism[1], and it is common in magnetic materials. The property of zero net moment that is inherent in antiferromagnets due to staggered spins has advantages such as the absence of a stray field and the feature of intrinsically fast spin dynamics[2-4]. However, antiferromagnetic (AFM) materials had limited utilization in spintronic applications and played an auxiliary role in conventional ferromagnetic (FM) spintronics until the new field of AFM spintronics was established[2,4-7]. The field of AFM spintronics aims to realize spintronic functional properties by utilizing the irreplaceable benefits of antiferromagnets. One such fascinating property is ultrafast AFM dynamics that has enhanced the writing speed to a terahertz scale, which surpasses the intrinsic gigahertz threshold in ferromagnets[8]. Contrarily, one significant challenge to host an antiferromagnet is the difficulty of controlling and detecting an AFM state due to its robustness against external magnetic fields.

Efficient means to manipulate an AFM state have been developed, for instance, electrical switching[9-13], optical excitation[14,15], AFM domain control[14,16-18] and heat-assisted magnetic recording[19,20]. The fundamental feature for detecting a controlled AFM memory state is the magnetocrystalline anisotropy. Thus, anisotropic magnetoresistance (AMR) has been adopted to detect the preferred AFM spin axis[2,12,14,20-25]. However, in many cases, complex stacking geometry with additional reference layers was required for unified spintronic functionality.

From recent observations in AFM spintronics that an AFM order dominates dynamic transport through the system[9,26,27], our notion has changed to considering AFM materials as promising candidates for the future generation of spintronic technology. Here, we offer a novel single-phase spintronic material, antiferromagnetic $Ca_{0.9}Sr_{0.1}Co_2As_2$ (CSCA), in which a multiple stacking of 1 nm scale spin-valve-like structure is intrinsically formed via the spin-flips, revealing the AMR and large anomalous Hall conductivity. An essential role of magnetocrystalline anisotropy for the observed spintronic functionality has been verified by combined experimental and theoretical work. Further, various spin states in the spin-valve-type operation are identified, increasing applicablility for AFM spintronics.

## RESULTS
### Spin-flip transition and magnetoresistance anisotropy

The CSCA belongs to the $ThCr_2Si_2$-type structure family that has been extensively investigated in the interest of versatile magnetic and electronic states[28,29]. These types of compounds have received particular attention for their various forms of superconductivity such as Fe-based high-temperature ($T$) superconductivity in doped $BaFe_2As_2$[30,31], heavy fermion superconductivity in $CeCu_2Si_2$[32,33], and other types, including various transition metals in $BaNi_2As_2$[34,35], $LaIr_2Ge_2$[36], and $LaRu_2P_2$[37]. The CSCA single crystals form a tetragonal structure (I4/mmm space group). Two $Co_2As_2$ layers lie opposite to each other around the center of a unit cell, separated by a non-magnetic (NM) Ca/Sr single layer, as depicted in Fig. 1a. This unit is stacked repeatedly along the $c$-axis. It has been known that Co spins, which are FM within a layer, are coupled antiferromagnetically to the spins in the neighboring layer, giving rise to complete cancellation of the magnetic moments, i.e., the A-type AFM order (Fig.1a)[28,29]. The structural units are elucidated in the scanning transmission electron microscopy (STEM) images (Fig.1b). Lattice constants were attained by fast Fourier transformation from the STEM data as $a$ = 0.408 nm and $c$ = 1.086 nm. The AFM order emerges at $T_N \approx$ 97 K, exhibited as an anomaly in the magnetic susceptibility and $T$ derivative of resistivity (see Supplementary Note 1 for details).

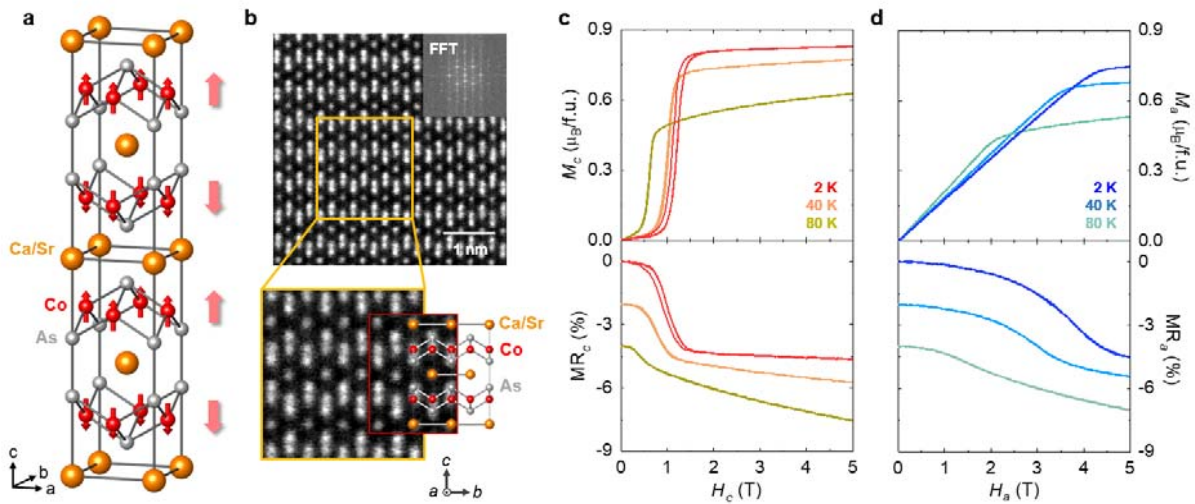

**Fig. 1 Structure and magnetic-field dependent magnetization and magnetoresistance. a** Crystal structure of body-centered tetragonal $Ca_{0.9}Sr_{0.1}Co_2As_2$. The orange, red and grey spheres represent Ca/Sr, Co, and As atoms, respectively. The red arrow on the Co atom denotes the individual spin direction. The right red arrow indicates the net magnetic moment of the $Co_2As_2$ layer. **b** Dark-field images in the $bc$ plane, taken using a scanning transmission electron microscope (STEM). The STEM image with lower magnification indicates that all the layers are regularly aligned. Inset shows the diffraction pattern obtained from fast Fourier transformation. **c** Isothermal magnetization and magnetoresistance along the $c$-axis at $T$ = 2, 40 and 80 K. To make it easier to distinguish, magnetoresistance data are shifted vertically. **d**

Isothermal magnetization and magnetoresistance along the *a*-axis at *T* = 2, 40 and 80 K.

Antiferromagnetism is generally explained by a collinear two-sublattice model[38,39]. In zero magnetic-field (*H*), two-sublattice magnetization (*M*) vectors with equal magnitude are aligned in opposite directions. A sufficient strength of *H* along the AFM spin axis may induce flops or flips of the *M* vectors, resulting in a magnetic transition. The main feature of this transition is the spin reorientation through phase conversion, which renders distinct anomalies in physical properties[40-43]. A large step-like increase in $M_c$ (*M* along the *c*-axis) at $H_{\text{flip}}$ = 1.2 T and *T* = 2 K results in a spin-flip transition as shown in Fig. 1c. A noticeable magnetic hysteresis can be identified, reflecting a first-order nature of the transition. Upon increasing *T*, the $H_{\text{flip}}$, determined by the *H*-derivative of $M_c$, is progressively reduced from 1.2 T at 2 K to 0.6 T at 80 K. Contrarily, in the $M_a$ (*M* along the *a*-axis) at 2 K, the gradual canting of Co spins leads to a linear increase up to 4.0 T, above which the slope of $M_a$ is considerably reduced (Fig. 1d). The *H* at which the slope of $M_a$ changes is lowered as *T* increases. To investigate the influence of anisotropic *M* on transport, we measured the magnetoresistance, MR = $\frac{R(H)-R(0)}{R(0)}$ for the *c*- and *a*-axes (Figs. 1c and 1d). The value of $\text{MR}_c$ also shows an abrupt drop at $H_{\text{flip}}$. However, $\text{MR}_a$ decreases monotonously upon increasing *H*. This close correspondence between *M* and MR plots suggests that magnetic order dominates the magneto-transport and its anisotropy.

**Large anomalous Hall conductivity and anisotropic magnetoresistance**
The electrical Hall effect can be largely improved via the interplay between conduction electrons and magnetism. In ferromagnets, an extra contribution to the ordinary Hall effect is shown to be proportional to *M* due to the spin-orbit coupling, which contribution is known as anomalous Hall effect (AHE). In non-collinear antiferromagnets, a large AHE was observed, despite the vanishingly small magnitude of *M*[44,45]. This AHE originates from the non-zero Berry curvature associated with topologically non-trivial spin textures[46,47]. In our collinear AFM CSCA with a strong magnetocrystalline anisotorpy, a certain *H* accompanies a spin-flip transition involving a drastic change in *M*. This spin-reorientation feature can be monitored by the AHE resulting from considerable spin-orbit coupling in the CSCA. As shown in Fig. 2a, after subtraction of small linear component originating from the ordinary Hall effect, the transverse conductivity, defined as $\sigma_{yx}^A = \rho_{yx}/(\rho_{xx}^2 + \rho_{yx}^2)$ directly follows the variation of *M*. The maximum $\sigma_{yx}^A$ of ~200 $\Omega^{-1}\text{cm}^{-1}$ with a maximum anomalous Hall angle, $\Theta_{\text{AH}}$ = $\Delta\sigma_{yx}^A/\sigma_{xx} \approx 2.24$ % was measured at 2 K, and it reduces to a half magnitude at 80 K. The $\sigma_{yx}^A$ can be well scaled by $M$ with the scaling factor $S_H = \sigma_{yx}^A/M \approx 0.214$ $V^{-1}$ at 2 K, and 0.156 $V^{-1}$ at 80 K. Figure 2b displays the contour plots of $\sigma_{yx}^A$ obtained from the angular dependence of $\sigma_{yx}^A$ at various $H$ values at 2 K, which clarifies the spin-flip driven emergence of the large AHE.

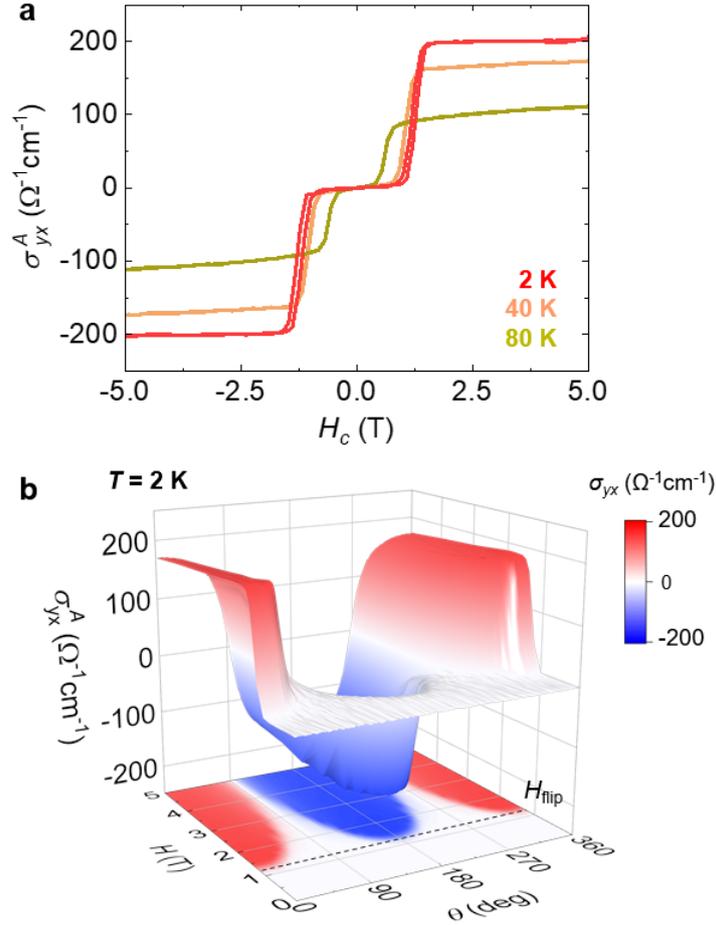

**Fig. 2 Large anomalous Hall effect. a** The $H$ dependence of anomalous Hall conductivity, $\sigma_{yx}^A$, for $T$ = 2, 40, and 80 K. **b** 2D and 3D contour plots of $\sigma_{yx}^A$, established from the angle dependence of $\sigma_{yx}^A$ at various $H$ values at $T$ = 2 K. Gray dotted line denotes the spin-flip transition, $H_{flip}$ = 1.2 T at 2 K.

By plotting $\Delta$MR = MR$_a$ – MR$_c$, a distinct MR difference can be demonstrated as shown in Fig. 3a. A steep variation develops in the spin-flip regime, and a maximal difference is noticeable just after the spin-flips at $H_{max} \approx 1.6$ T for $T$ = 2 K. Similar features accompanying the reduced $\Delta$MR and lowered $H_{max}$ were observed at $T$ = 40 and 80 K. The AMR, defined as $\frac{R(\theta)-R(0)}{R(0)}$, was measured with the geometry shown in Fig. 3b. Note that the $H$ is continually applied perpendicular to the current, inhibiting the ordinary Lorentzian MR effect. The uniaxial anisotropy allows two-fold rotational symmetry, which is presented in the polar angular plot of the AMR (Fig.3b). The AMR reveals a dumbbell-like shape in which the maximum value occurs at $\theta$ = 90 and 270°. Across the $H_{flip}$, the AMR is considerably enhanced, and the largest variation occurs at $H_{max}$ = 1.6 T, which is consistent with ΔMR. The maximum AMR at $H_{max}$ reaches ~4 %, one order larger than those of other AFM metals[19,20,48]. The complete AMR contour map in Fig. 3c reveals that the AMR effect is obviously maximized at $H_{max}$. Above $H_{max}$, the AMR is gradually reduced because of weakened anisotropy. The decrease in $H_{max}$ is displayed in the contour plots at $T$ = 40 and 80 K (Fig. 3d), coinciding with the ΔMR behavior (Fig. 3a). Detailed $T$ evolution of the AMR effect is shown in the $\theta$-$T$ contour plots in Supplementary Note 2.

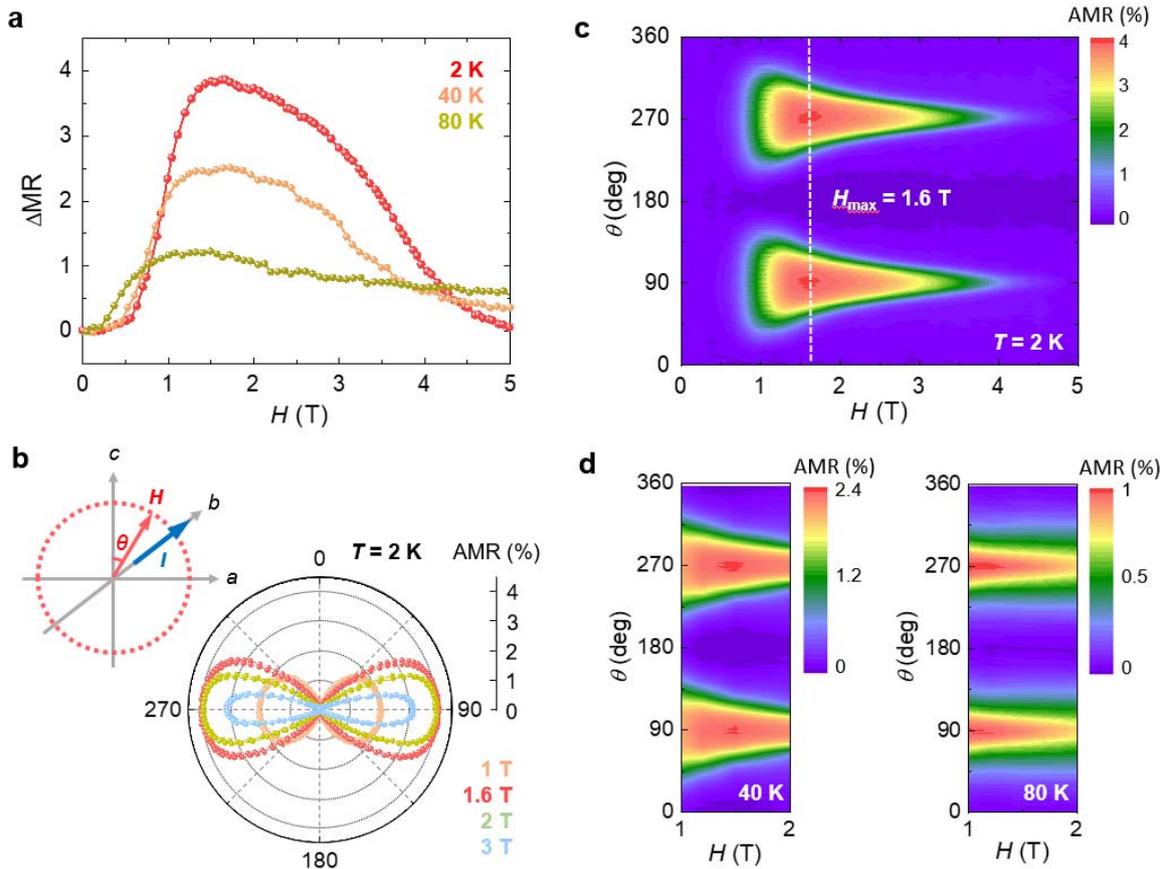

**Fig. 3 Anisotropic magnetoresistance. a** The MR difference between $a$- and $c$-axes, ΔMR = $MR_a$ − $MR_c$, taken at $T$ = 2, 40 and 80 K. **b** Polar angular plot of the anisotropic magnetoresistance (AMR), measured at $T$ = 2 K by rotating $H$ =1, 1.6, 2 and 3 T in the $ac$ plane with the current along the $b$-axis, $I//b$. Geometry of the AMR measurement is schematically shown. $\theta$ = 0° for the $c$-axis and $\theta$ = 90° for the $a$-axis. **c** Contour plot of the AMR measured

at $T = 2$ K. **d** Contour plots of the AMR taken at $T = 40$ and 80 K.

**Multiple stacking of 1-nm-scale spin-valve-type units driven by Spin-flips**

To clarify the nature of the magnetic phase transition, theoretical calculations based on the spin Hamiltonian with uniaxial magnetocrystalline anisotropy were performed. By fitting the theoretical results of anisotropic magnetizations to the experimental data, we estimate the parameters $J$ and $K$: $g\mu_B H_{\text{flip}}/JS = 2$, where $S$ is the saturation magnetic moment and $K = 1.4\ JS^2$. This belongs to the strong magnetocrystalline anisotropy, as $K > JS^2$ for our model Hamiltonian. However, experimentally at the $H$-induced transition in the CSCA, the step-like increase does not reach the saturated state for two reasons. First, a non-linear regime appears just after the large step-like increase (Fig. 1c) due to spatial modulations of phase coexistence regarding the first-order nature of the transition[49]. By considering the spatial modulations, we present the resulting $M_c$ and $M_a$ in Figs. 4a and 4c, compatible with the experimental data (Figs. 1c and 1d). Second, a slight linear slope was observed above the nonlinear regime. This slope increases continually as $T$ rises with the reduction in $M_c$ value at the highest measured $H$ (Fig.1c), suggesting that this slope is attributed to the thermal fluctuation inhibiting saturation of the $M_c$.

A spin-flip transition can be understood by the essential influence from strong magnetocrystalline anisotropy, found in the magnetic response of the net magnetic moments. A schematic layered structure in Fig. 4a represents two FM Co layers with relatively oriented net moments that are split by a NM Ca/Sr layer in a unit cell. The magnetocrystalline anisotropy tends to bind spin directions to a specific crystal axis. As a result, the AFM phase of CSCA in zero $H$ is stabilized along the $c$-axis, with the total magnetic energy minimized. The $H$ along the $c$-axis ($H_c$) activates the spin-flips, accompanied by a large increase in $M_c$, i.e., flips of net moments. By contrast, $H_a$ generates steadily canted net moments, which signifies the presence of a highly anisotropic nature (Fig. 4c).

Closely looking at the process of the flip transition, the up-down net moment state describing the AFM phase evolves to the up-up net moment state, which suggests that this structure is largely akin to a simple spin-valve- or a giant magnetoresistance-type (GMR-type) device that is artificially constructed by magnetic layers, as depicted in Fig. 4b. Spin-valve devices have been employed for microelectronic applications such as magnetic-field sensors[50], read heads in a hard drive[51] and magnetic random access memories[52]. In our CSCA, repeated stacking of the atomic-scale multilayered unit approximately 1 nm thick ($c$ = 1.086 nm) constitutes a multiple spin-valve-type structure.

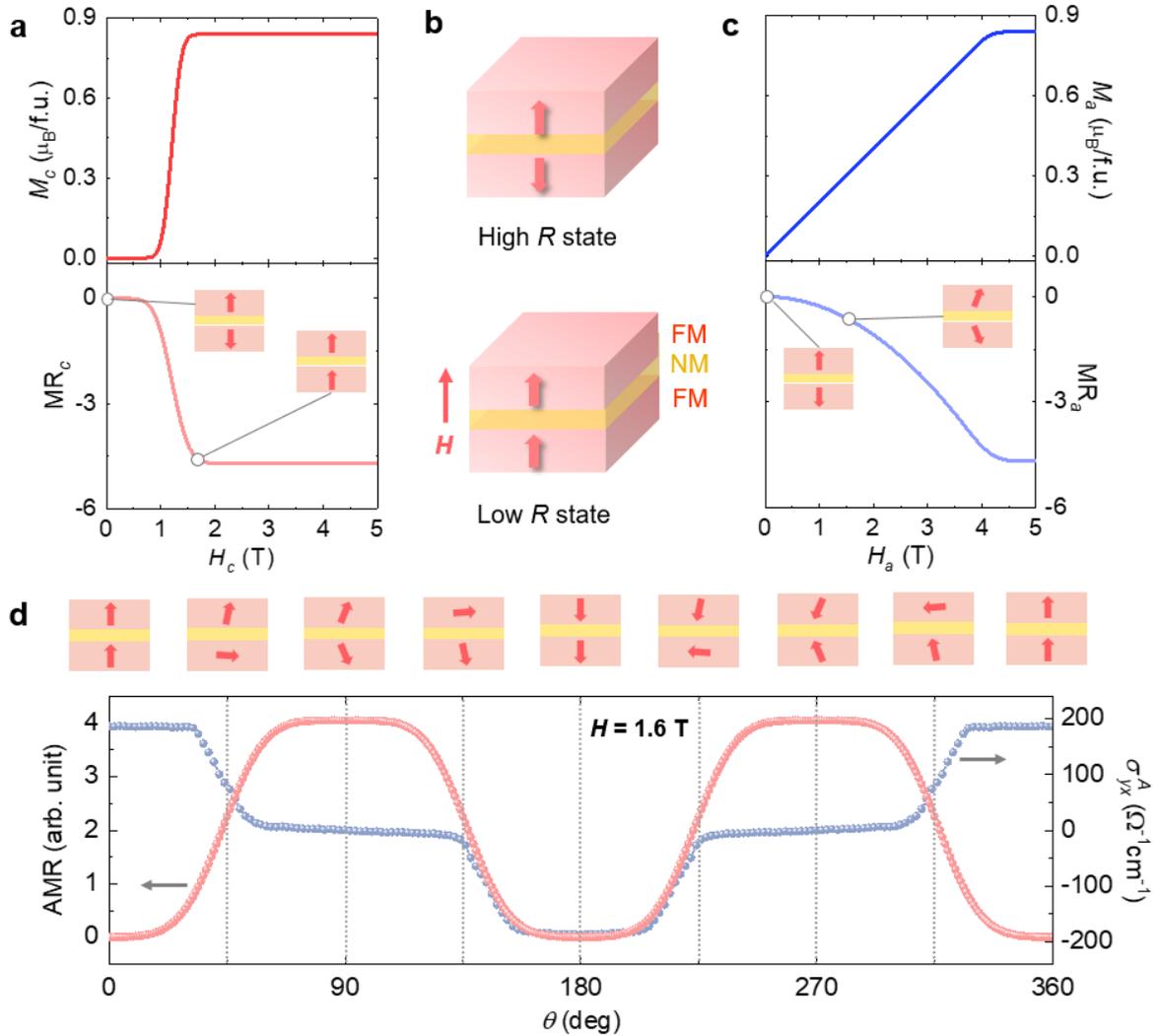

**Fig. 4 Intrinsic spin-valve-type structure and electrically-detectable spin states. a** Calculated isothermal $M_c$ and MR$_c$. A schematic layered structure presents two ferromagnetic (FM) Co layers with relatively oriented net moments in a specific $H$, separated by a nonmagnetic (NM) Ca/Sr layer in a unit cell. The thickness of the naturally formed spin-valve-type unit is approximately 1 nm ($c$ = 1.086 nm). **b** Schematic low and high resistance ($R$) states of bilayer GMR-type spin-valve structure with out-of-plane magnetizations. Two FM layers are separated by a NM spacer. The arrows within each FM layer denote the direction of net $M$. **c** Calculated isothermal $M_a$ and MR$_a$. **d** Switching operation between high and low $R$ states of spin-valve-like structure by rotating $H$ in the $ac$ plane. The AMR was obtained by conductance calculation for $H_{max}$ = 1.6 T. The angular dependence of $\sigma_{yx}^A$ was measured at 2 K for $H_{max}$. Schematics present relative orientations of net magnetic moments in a unit cell with respect to angle variation of *H*, estimated by easy-axis magntocrystalline anisotropy model.

**Spin-valve-like operation by a rotating magnetic field and electrical detection of spin states**

To theoretically calculate the magneto-transport property, we propose that the interlayer hopping amplitude is given by $t_{i,i+1} = |\langle \hat{n}_i | \hat{n}_{i+1} \rangle|$ ($i$ = 1, 2). Here $|\langle \hat{n}_i | \hat{n}_{i+1} \rangle|$ denotes the overlap integral between two spinors, where each spinor is aligned parallel to the spins in each layer. It is given by $\cos\frac{\gamma}{2}$, where $\gamma$ represents the relative angle between two spinors. For the MR calculations, we assume that the conductance ($\sigma$) of the system is proportional to the multiple hopping amplitudes through the bilayer system, and hence, $\sigma \propto t_{12}t_{23}$. From the definition of MR and $R = 1/\sigma$, MR is proportional to $\sigma(0) - \sigma(H)$. The development of anisotropic magneto-transport properties through the flip transition can be reasonably described by the $\sigma$ calculated directly from relative spin orientations between different $Co_2As_2$ layers as shown in Figs. 4a and 4c. The measured and calculated MR$_c$ in Figs. 1c and 4a, respectively, unambiguously presents the switchable low/high resistance state. This new type of spin-flip-driven mechanism is distinguished from that of a conventional spin-valve device. The synthetic antiferromagnets[53] manufactured by nano-fabrication methods derive the benefits of natural antiferromagnets, such as hosting fast domain wall motion[54] and peculiar spin texture in the form of skyrmions[55]. However, the experimental achievement of spintronic functionality in synthetic magnetic multilayers is greatly susceptible to defect/dislocation, interfacial roughness, and impeccable epitaxy[23,56,57]. Our simple-resistor-type CSCA is advantageous in that a spin-valve-like arrangement is intrinsically embedded in the single-crystalline AFM structure without such interfacial flaws.

In the presence of a rotating *H*, a unique type of the spin-valve-like operation was also observed, based on the peculiar AMR and AHE characteristics. Conduction electrons are crucial to track the various spin states formed during the switching operation by rotating $H_{max}$ = 1.6 T for the multiple spin-valve-type structure. Figure 4d shows the AMR and angular dependence of AHE at $H_{max}$. Combined with theoretical calculations, the different memory states are electrically probed as shown in the schematics. The various spin states are clearly discernible for the AMR and angular dependence of AHE at $H_{max}$. With the rotation of *H* from 0 to 90°, the net moments that are parallel convert to the antiparallel arrangement by switching the net moment in one of the two layers, which transforms from a low to a high *R* state or from positive to zero $\sigma_{yx}^A$ state.

Further rotation from 90 to 180° induces the high to low $R$ or zero to negative $\sigma_{yx}^A$ switching by orienting the net moment in the other of the two layers. The AHE apparently satisfies Onsager relation of $\sigma_{yx}^A(\vec{H}) = -\sigma_{yx}^A(-\vec{H})$.

**DISCUSSION**

We have demonstrated improved anisotropy of magnetic properties, large AHE, and AMR effect via the spin-flip transition in a single-phase antiferromagnet. Based on our results, to detect clear functional properties, two conditions need to be fulfilled. One is the emergence of a well-defined and complete magnetic transition, across which the spin-valve-type multiple structures on an atomic scale are naturally formed within an A-type AFM phase. In the $Ca_{1-x}Sr_xCo_2As_2$ compounds, interlayer magnetic couplings and magneto crystalline anisotropy seem susceptible to the distance between magnetic layers, which can be controlled by chemical doping[28]. The 10% Sr-doped compound ($x = 0.1$), which is the compound considered in our study (CSCA), was selected because the magnetic transition is optimized with a much sharper and more intact step-like shape, compared to that of $CaCo_2As_2$[29,58]. In addition, the $H_{flip}$ is lowered with increasing $T_N$[29]. After further Sr-doping ($x = 0.2$), the modified interlayer coupling leads to a complete phase change resulting in a FM state[28]. The other is the large anisotropic nature, which maximizes the difference between MR curves for different orientations. Basically, to stabilize the AFM arrangement, a sufficient degree of magnetocrystalline anisotropy is required for a given state fixed in one direction. Further, the flip transition generates a drastic change in anisotropy of magnetic properties. Therefore, the combined effect of atomic-scale spin-valve-like structure and large anisotropic nature is responsible for the AFM spintronic functionality in a natural antiferromagnet of CSCA.

In summary, we suggest a new single-crystalline spintronic antiferromagnet, $Ca_{0.9}Sr_{0.1}Co_2As_2$, in which a multiple structure composed of 1 nm scale spin-valve-type units is inherently introduced through the spin-flip transition. This intrinsic structure reveals the spin-flip-driven large anomalous Hall conductivity. The switching operation between low and high resistance states was demonstrated by rotating a magnetic field. This leads to anisotropic magnetoresistance, which is maximized in the vicinity of the flip transition. Our theoretical estimation verifies that the easy-axis magnetocrystalline anisotropy is a key factor for the observed spintronic functionality. Electrical detection for the diverse memory states is applicable to a wide range of antiferromagnets. The spintronic functionality using intrinsic bulk properties simplifies the stacking geometry, thereby rendering a platform to develop functional heterostructures.

## METHODS

### Sample preparation

The $Ca_{0.9}Sr_{0.1}Co_2As_2$ single crystals were grown by the self-flux method[29,59]. A CoAs precursor was first prepared by the solid state reaction using mixed powders of Co (99.5%, Alfa Aesar) and As (99,999%, Sigma Aldrich) with a fixed stoichiometric ratio and calcined in air at 700 °C for 24 h in a furnace. The CoAs powder was blended with Ca and Sr flakes at a 4:0.9:0.1 molar ratio of CoAs:Ca:Sr, and the mixture was placed in an alumina crucible. The crucible was vacuum-sealed in a quartz tube. In a high-temperature furnace, the quartz tube was dwelled at 1280 °C for 16 h, slowly cooled to 850 °C at a rate of 2 °C/h, and then cooled to room temperature at a rate of 100 °C/h. Crystals with typical dimensions of 1.5 × 3 × 0.2 mm$^3$ were obtained.

### Scanning transmission electron microscopy (STEM) measurement

Samples for STEM were prepared with a cutting plane perpendicular to the *a*-axis using a dual-beam focused ion beam system (Helios 650, FEI). The cutting plane displays a well-identifiable atomic structure. To avoid critical damage on a thin sample, acceleration voltage conditions were varied from 30 to 2 keV. The atomic structure was obtained utilizing STEM (JEM-ARM200F, JEOL Ltd, Japan) at 200 keV with a Cs-corrector (CESCOR, CEOS GmbH, Germany) and cold field emission gun. The size of the electron probe was 83 pm, and the range of high-angle annular dark-field detector angle was in the range of 90 to 370 mrad.

### Magnetic and transport property measurements

Magnetic susceptibility and isothermal magnetization measurements were carried out with magnetic fields along *a*- and *c*-axes using a vibrating sample magnetometer (VSM) module in a physical property measurement system (PPMS, Quantum Design, Inc.). The magnetic-field dependences of in-plane resistivity and Hall resistivity measurements were performed using the conventional six-probe configuration in the PPMS. The anisotropic magnetoresistance and angle dependence of anomalous Hall resistivity were obtained by polar angle scan of magnetic fields in the *ac* plane in the PPMS equipped with a single-axis rotator.

### Easy-axis anisotropic spin model

The spin Hamiltonian with uniaxial magnetocrystalline anisotropy can be expressed as

$$\mathcal{H}/N = J\sum_{i=1}^{2} \vec{S}_i \cdot \vec{S}_{i+1} - g\mu_B \vec{H} \cdot \sum_{i=1}^{2} \vec{S}_i + K\sum_{i=1}^{2} \sin^2\theta_i,$$

where $J$ represents the AFM coupling strength between Co moments in adjacent layers, $N$ denotes the number of Co moments in a single layer, $g = 2$, and $K$ denotes the magnetocrystalline anisotropy constant. The second term represents the Zeeman energy, where the magnetic field $\vec{H}$ lies on the *ac* plane making an angle $\theta$ with the *c*-axis. The third term denotes the uniaxial magnetocrystalline anisotropy energy, which is consistent with the favorable spin orientation along the *c*-axis.

## DATA AVAILABILITY

The data that support the findings of this study are available from the corresponding authors upon request.

**ACKNOWLEDGEMENTS**

The work was supported by the National Research Foundation of Korea (grant numbers NRF-2016R1D1A1B01013756, NRF-2017R1A5A1014862 (SRC program: vdWMRC center), NRF-2019R1A2C2002601, and NRF-2021R1A2C1006375).

**AUTHOR CONTRIBUTIONS**

N.L. and Y.J.C. conceived the project. D.G.O. and K.W.J. synthesized the single crystals. D.G.O., J.H.K., K.W.J., H.J.S., J.M.H., and J.S.K. performed measurements of the physical properties of the crystals. M.K.K. and K.M. performed the theoretical calculations. D.G.O., J.H.K., K.M., N.L. and Y.J.C. analyzed the data and prepared the manuscript. All the authors have read and approved the final version of the manuscript.

**COMPETING INTERESTS**

The authors declare no competing interests.

**ADDITIONAL INFORMATION**

**Supplementary information** The online version contains supplementary material.

**Correspondence** and requests for materials should be addressed to Nara Lee or Young Jai Choi.

**Reprints and permission information** is available at http://www.nature.com/reprints

**Publisher's note** Springer Nature remains neutral with regard to jurisdictional claims in published maps and institutional affiliations.

# Supplementary Information

# Large anomalous Hall effect and anisotropic magnetoresistance in intrinsic nanoscale spin-valve-type structure of an antiferromagnet

Dong Gun Oh[1,2], Jong Hyuk Kim[1,2], Mi Kyung Kim[1], Ki Won Jeong[1], Hyun Jun Shin[1], Jae Min Hong[1], Jin Seok Kim[1], Kyungsun Moon[1], Nara Lee[1,*], and Young Jai Choi[1,*]

[1]Department of Physics, Yonsei University, Seoul 03722, Korea. [2]These authors contributed equally: D. G. Oh, J. H. Kim. *email: eland@yonsei.ac.kr; phylove@yonsei.ac.kr

**This PDF file includes:**

Supplementary Note 1. Temperature dependence of antiferromagnetic properties in $Ca_{0.9}Sr_{0.1}Co_2As_2$ crystals

Supplementary Note 2. Temperature evolution of anisotropic magnetoresistance

**Supplementary Note 1. Temperature dependence of antiferromagnetic properties in $Ca_{0.9}Sr_{0.1}Co_2As_2$ crystals**

Magnetic susceptibility, $\chi = M/H$, measured at $H = 0.1$ T upon warming after zero-field-cooling, indicates the emergence of antiferromagnetic (AFM) order at $T_N \approx 97$ K, as plotted in Supplementary Fig. 1a. Magnetic measurements were carried out with magnetic fields along $a$- and $c$-axes using a vibrating sample magnetometer (VSM) module in a physical property measurement system (PPMS, Quantum Design, Inc.). The anisotropic nature has manifested in the $\chi$ curves for the $H$ along the $a$- and $c$-axes. The faster decrease of $\chi$ below $T_N$ along the $c$-axis is consistent with the magnetic moments of Co ions aligning along this axis. Electric transport measurements were carried out at zero $H$ in the PPMS. The resistivity as a function of

$T$ exhibits a metallic behavior and a distinct anomaly at $T_N$ identified in its $T$ derivative (Supplementary Fig. 1b).

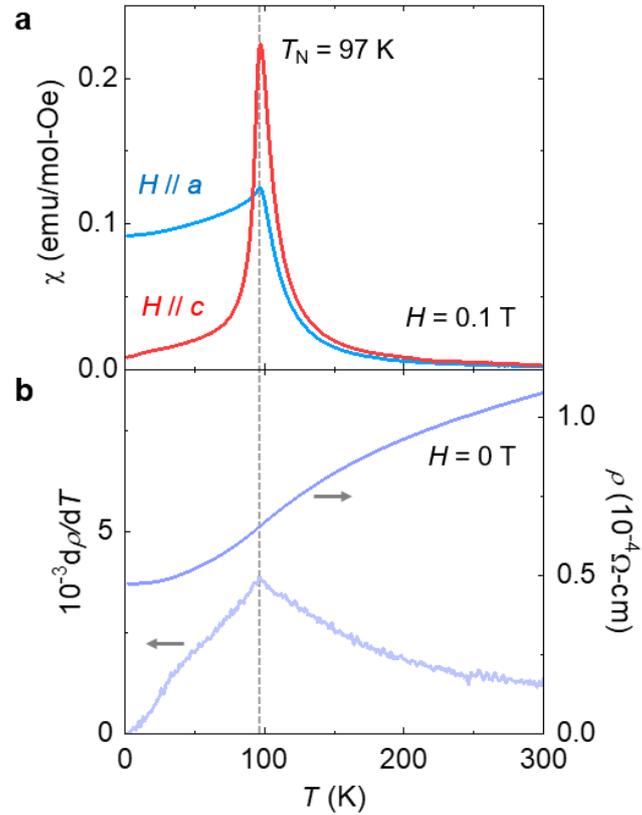

**Supplementary Fig 1 Temperature dependence of antiferromagnetic properties. a** Temperature dependence of the magnetic susceptibility $\chi = M/H$, measured upon warming at $H$ = 0.1 T after zero-field cooling for the $a$- and $c$-axes. Vertical grey line indicates the Néel temperature, $T_N$ = 97 K. **b** Temperature dependence of resistivity at zero magnetic field and its temperature derivative.

**Supplementary Note 2. Temperature evolution of anisotropic magnetoresistance**

We have examined the $T$ evolution of the AMR = $\frac{R(\theta)-R(0)}{R(0)}$, measured in the geometry shown in Supplementary Fig. 2a. The AMR comprises two different components. A noncrystalline component may arise from the relative orientation between $M$ and a specific current direction, and a crystalline component is relevant entirely to crystal symmetry[1-3]. A conventional AMR effect in polycrystalline 3$d$ ferromagnetic alloys was ascribed predominantly to the noncrystalline AMR component as all crystalline components average out. In pursuance of the AMR effect driven dominantly by magnetocrystalline anisotropy in AFM spintronics, the crystalline AMR component for single-crystalline bulk and thin film materials has particularly been demonstrated[4,5]. The two-fold rotational symmetry is reflected in the polar angular plot of the AMR measured at $H_{max}$ = 1.6 T (Supplementary Fig. 2a). The dumbbell-like shape indicating the uniaxial magnetocrystalline anisotropy diminishes progressively as $T$ is increased to 80 K. The detailed $T$ variation of the AMR is plotted in the $\theta$-$T$ contour plot (Supplementary Fig. 2b). The smaller magnitude but broader shape of the AMR was found at $H$ = 1 T (Supplementary Fig. 2c). At $H$ = 3 T, the narrower shape of the AMR almost disappears around 60 K (Supplementary Fig. 2d).

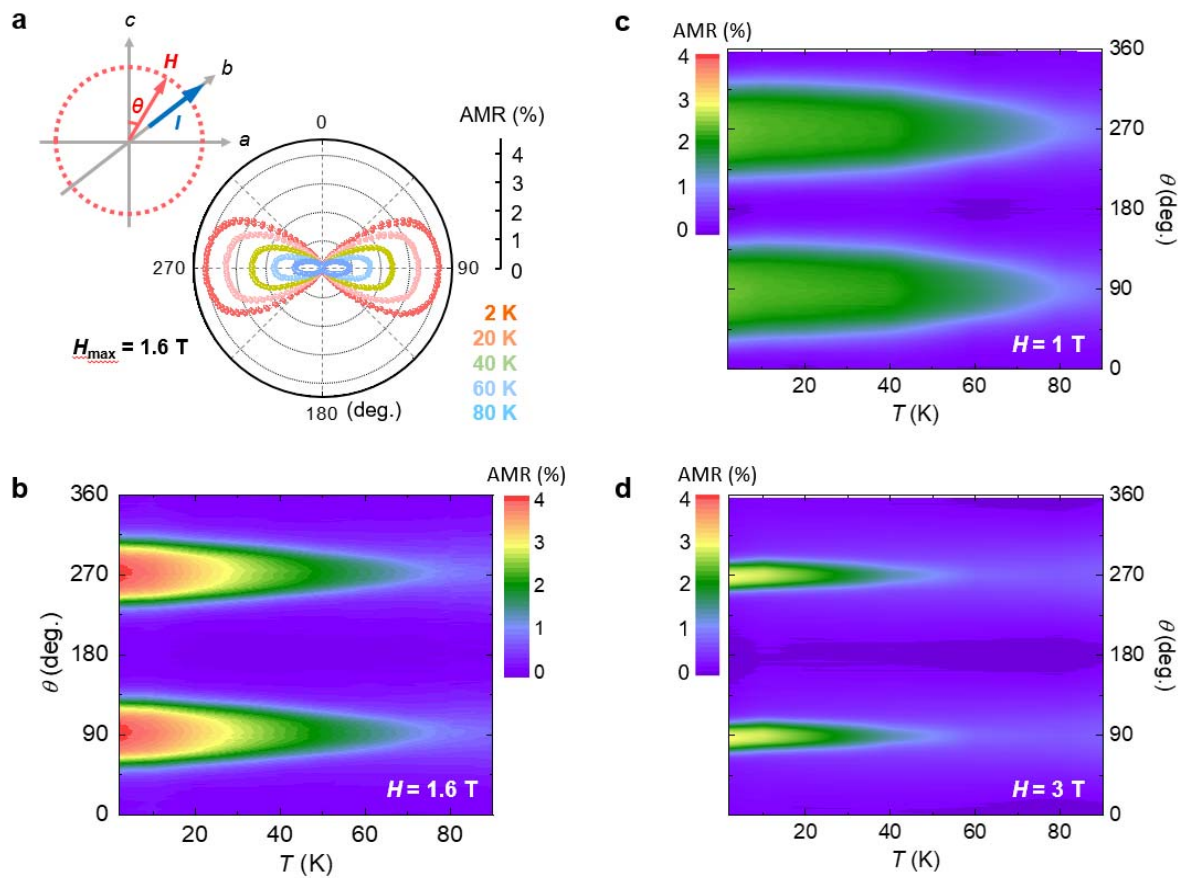

**Supplementary Fig 2 Temperature evolution of anisotropic magnetoresistance**. **a** Geometry of the AMR measurement is schematically shown. $H$ is rotated in the $ac$ plane, while the current is applied along the $b$-axis, $I//b$. Polar angular plot of the AMR, measured at $T$ = 2,

20, 40, 60, and 80 K by rotating $H_{max}$ =1.6 T. **b** $\theta$-$T$ contour plot constructed from the AMR data measured at various values of $T$ and $H_{max}$ =1.6 T. **c-d** $\theta$-$T$ contour plots of the AMR for $H$ =1 and 3 T, respectively.